\documentclass[dvipdfmx]{ws-ijmpa}
\usepackage[super,compress]{cite}
\usepackage{graphicx}
\usepackage{braket}
\usepackage{bm}
\usepackage{comment}
\usepackage{here}

\begin{document}
\markboth{Yuta Hyodo}{Magic square and Dirac flavor neutrino mass matrix}

\def\Journal#1#2#3#4{{#1} {\bf #2}, #3 (#4)}
\def\AHEP{Advances in High Energy Physics.} 
\def\ARNPS{Annu. Rev. Nucl. Part. Sci.} 
\def\AandA{Astron. Astrophys.} 
\def\ANP{Ann. Phys.}
\def\APJ{Astrophys. J.}
\def\APJS{Astrophys. J. Suppl}
\def\COMR{Comptes Rendues}
\def\CPC{Chin. Phys. C}
\def\EPJC{Eur. Phys. J. C}
\def\EPL{EPL}
\def\IJMPA{Int. J. Mod. Phys. A}
\def\IJMPE{Int. J. Mod. Phys. E}
\def\JCAP{J. Cosmol. Astropart. Phys.}
\def\JHEP{J. High Energy Phys.}
\def\JETPL{JETP. Lett.}
\def\JETPUSSR{JETP (USSR)}
\def\JPG{J. Phys. G} 
\def\JPGNP{J. Phys. G: Nucl. Part. Phys.} 
\def\MPLA{Mod. Phys. Lett. A}
\def\NIMA{Nucl. Instrum. Meth. A.}
\def\NATU{Nature}
\def\NCA{Nuovo Cimento}
\def\NJP{New. J. Phys.}
\def\NPB{Nucl. Phys. B}
\def\NPBOLD{Nucl. Phys.}
\def\NPBSUPPL{Nucl. Phys. B. Proc. Suppl.}
\def\PL{Phys. Lett.}
\def\PLB{{Phys. Lett.} B}
\def\PMCA{PMC Phys. A}
\def\PREP{Phys. Rep.}
\def\PPNP{Prog. Part. Nucl. Phys.}
\def\PLBOLD{Phys. Lett.}
\def\PAN{Phys. Atom. Nucl.}
\def\PRL{Phys. Rev. Lett.}
\def\PRD{Phys. Rev. D}
\def\PRC{Phys. Rev. C}
\def\PR{Phys. Rev.}
\def\PTP{Prog. Theor. Phys.}
\def\PTEP{Prog. Theor. Exp. Phys.}
\def\RMP{Rev. Mod. Phys.}
\def\RPP{Rep. Prog. Phys.}
\def\SJNP{Sov. J. Nucl. Phys.}
\def\SCIENCE{Science}
\def\TNYAS{Trans. New York Acad. Sci.}
\def\ZETP{Zh. Eksp. Teor. Piz.}
\def\ZFPH{Z. fur Physik}
\def\ZPC{Z. Phys. C}

%
\catchline{}{}{}{}{}
%


\title{Magic square and Dirac flavor neutrino mass matrix}

\author{Yuta Hyodo}

\address{Course of Physics, Graduate School of Science, Tokai University,\\
4-1-1 Kitakaname, Hiratsuka, Kanagawa 259-1292, Japan\\
0CSNM017@mail.u-tokai.ac.jp}

\author{Teruyuki Kitabayashi}

\address{Department of Physics, Tokai University,\\
4-1-1 Kitakaname, Hiratsuka, Kanagawa 259-1292, Japan\\
teruyuki@tokai-u.jp}

\maketitle

\begin{history}
\received{Day Month Year}
\revised{Day Month Year}
\end{history}

\begin{abstract}
The magic texture is one of the successful textures of the flavor neutrino mass matrix for the Majorana type neutrinos. The name ``magic" is inspired by the nature of the magic square. We estimate the compatibility of the magic square with the Dirac, instead of the Majorana, flavor neutrino mass matrix. It turned out that some parts of the nature of the magic square are appeared approximately in the Dirac flavor neutrino mass matrix and the magic squares prefer the normal mass ordering rather than the inverted mass ordering for the Dirac neutrinos.
\end{abstract}

\ccode{PACS numbers:14.60.Pq}


\section{Introduction\label{section:introduction}}
Understanding the nature of the flavor structure of elementary particles is one of the long-term problems in particle physics and cosmology \cite{King2015JPG,Feruglio2019}. Many texture ansatz to solve the flavor puzzle are proposed, such as tri-bi maximal texture \cite{Harrison2002PLB,Xing2002PLB,Harrison2002PLB2}, texture zeros \cite{Berger2001PRD,Frampton2002PLB,Xing2002PLB530,Xing2002PLB539,Kageyama2002PLB,Xing2004PRD,Grimus2004EPJC,Low2004PRD,Low2005PRD,Grimus2005JPG,Dev2007PRD,Xing2009PLB,Fritzsch2011JHEP,Kumar2011PRD,Dev2011PLB,Araki2012JHEP,Ludle2012NPB,Lashin2012PRD,Deepthi2012EPJC,Meloni2013NPB,Meloni2014PRD,Dev2014PRD,Felipe2014NPB,Ludl2014JHEP,Cebola2015PRD,Gautam2015PRD,Dev2015EPJC,Kitabayashi2016PRD,Zhou2016CPC,Singh2016PTEP,Bora2017PRD,Barreiros2018PRD,Barreiros2019JHEP,Capozzi2020PRD,Singh2020EPL,Barreiros2020}, $\mu -\tau$ symmetric texture \cite{Fukuyama1997,Lam2001PLB,Ma2001PRL,Balaji2001PLB,Koide2002PRD,Kitabayashi2003PRD,Koide2004PRD,Aizawa2004PRD,Ghosal2004MPLA,Mohapatra2005PRD,Koide2005PLB,Kitabayashi2005PLB,Haba206PRD,Xing2006PLB,Ahn2006PRD,Joshipura2008EPJC,Gomez-Izquierdo2010PRD,He2001PRD,He2012PRD,Gomez-Izquierdo2017EPJC,Fukuyama2017PTEP} and textures under $A_n$ as well as $S_n$ symmetries \cite{Altarelli2010PMP}. 

So-called {\it magic texture} is one of the textures of the flavor neutrino mass matrix for Majorana type neutrinos \cite{Harrison2004PLB, Lam2006PLB}. The magic texture is parametrized as 
\begin{eqnarray}
M =\left(
\begin{matrix}
a & b & c \\
b & d & a+c-d \\
c & a+c-d & b-c+d
\end{matrix}
\right).
\label{Eq:magicTexture}
\end{eqnarray}
The applications of the magic texture have been studied for texture zeros of flavor neutrino mass matrix \cite{Gautam2016PRD}, with two simple extensions \cite{Channey2019JPGNP} and for baryon asymmetry of the Universe \cite{Verma2019arXiv}. 

The name ``magic" is inspired by the nature of the {\it magic square} \cite{Levitin2011}. A magic square of order $n$ is a $n \times n$ square grid filled with distinct natural numbers \cite{}. Each cell contains a number in the range $1,2,\cdots,n^2$. The sum of the numbers in each row, each column and diagonal is equal. For example, a magic square of order 3 is schematically shown as
\begin{eqnarray}
\begin{array}{|c|c|c|}
\hline
 \ 2 \ & \  7 \  & \ 6 \ \\
\hline
 9 & 5 & 1  \\
\hline
 4 & 3 & 8 \\  
\hline
\end{array}
\begin{array}{c}
\leftarrow 15 \\
\leftarrow 15 \\
\leftarrow 15 \\
\end{array}
\nonumber \\
\begin{array}{ccccccc}
& \nearrow & ~\uparrow & ~\uparrow & ~\uparrow & \nwarrow &\\
15 & & ~15 & ~ 15 & ~15 & &15  \\
\end{array} 
\end{eqnarray}
where the sum (which is called magic constant or magic sum) is fifteen. The magic texture in Eq.(\ref{Eq:magicTexture}) has a part of the nature of magic square, e.g., the sum of the elements in each row and each column is equal to $a+b+c$.

In this paper, we estimate the compatibility of the magic square with the Dirac, instead of the Majorana, flavor neutrino mass matrix by numerical calculations. We show that some parts of the nature of the magic square are appeared approximately in the Dirac flavor neutrino mass matrix. Moreover, as the main conclusion of this paper, we demonstrate that the magic squares prefer the normal mass ordering rather than the inverted mass ordering for the Dirac neutrinos.

This paper is organized as follows. In Section.\ref{section:classification}, ten types of magic square for the Dirac neutrinos are defined.  In Section.\ref{section:mass_matrix}, the magic nature of the Dirac flavor neutrino mass matrix is estimated. Section \ref{section:summary} is devoted to a summary.

\section{Classification \label{section:classification}}
For a matrix
\begin{eqnarray}
&& \qquad \quad
\left(
\begin{matrix}
a & ~b & ~c \\
d & ~e & ~f \\
g & ~h & ~i
\end{matrix}
\right)
\begin{array}{c}
\leftarrow S_1 \\
\leftarrow S_2 \\
\leftarrow S_3 \\
\end{array}
\nonumber \\
&&\begin{array}{ccccccc}
& \nearrow & \uparrow & \uparrow & \uparrow & \nwarrow &\\
~ ~T' & & S_4 & S_5 & S_6 & &T \\
\end{array} 
\label{Eq:magic_square}
\end{eqnarray}
we define the following eight sums
\begin{eqnarray}
&& S_1=a+b+c, \quad S_2=d+e+f, \quad S_3=g+h+i, \nonumber \\
&& S_4=a+d+g, \quad S_5=b+e+h, \quad S_6=c+f+i,  \nonumber \\
&& T=a+e+i, \quad T'=c+e+g,
\end{eqnarray}
and the following ten types of magic square for the Dirac neutrinos in the from of the $3 \times 3$ matrix.

\

{\bf Type-I:}
The exact magic square (we call it type-I magic square) should satisfy the following requirement
\begin{eqnarray}
S_1= S_2 = S_3 = S_4 = S_5 =S_6 = T = T'.
\label{Eq:const_type-I}
\end{eqnarray}
In this case, the standard deviation should vanish:
\begin{eqnarray}
sd_{\rm I} = \sqrt{\frac{\sum_{i=1}^6 (S_i-\bar{S})^2+(T-\bar{S})^2+(T'-\bar{S})^2}{8}},
\end{eqnarray}
where 
\begin{eqnarray}
\bar{S} = (S_1+ S_2 + S_3 + S_4 + S_5 +S_6 + T + T')/8,
\end{eqnarray}
is the average of the sums $S_1,S_2,\cdots,S_6,T$ and $T'$. For example, in the case of order 3 magic square
\begin{eqnarray}
\left( 
\begin{array}{ccc}
2 & 7  & 6  \\
9 & 5  & 1  \\
4 & 3 & 8  \\
\end{array}
\right),
\end{eqnarray}
the sums 
\begin{eqnarray}
S_1= S_2 = S_3 = S_4 = S_5 =S_6 = T = T'=15,
\end{eqnarray}
and the average $\bar{S}=15$ are obtained and the standard deviation becomes $sd_{\rm I}=0$. Moreover, we obtain the vanishing standard deviation for the following decimal magic square:
\begin{eqnarray}
\left( 
\begin{array}{ccc}
0.2 & 0.7  & 0.6  \\
0.9 & 0.5  & 0.1  \\
0.4 & 0.3 & 0.8  \\
\end{array}
\right).
\end{eqnarray}
Indeed, the sums of this decimal magic square
\begin{eqnarray}
S_1= S_2 = S_3 = S_4 = S_5 =S_6 = T = T'=1.5,
\end{eqnarray}
and the average $\bar{S}=1.5$ are obtained and the standard deviation becomes $sd_{\rm I}=0$.

The standard deviation $sd_{\rm I}$ is a good number to show that whether a matrix has the nature of the exact type-I magic square or not; however, the standard deviation is not good number for our purpose in this paper. For example, the matrix
\begin{eqnarray}
\left( 
\begin{array}{ccc}
2+1 & 7  & 6  \\
9 & 5  & 1  \\
4 & 3 & 8  \\
\end{array}
\right),
\label{Eq:type-I-ex1}
\end{eqnarray}
and
\begin{eqnarray}
\left( 
\begin{array}{ccc}
0.2+0.1 & 0.7  & 0.6  \\
0.9 & 0.5  & 0.1  \\
0.4 & 0.3 & 0.8  \\
\end{array}
\right),
\label{Eq:type-I-ex2}
\end{eqnarray}
have a small perturbation for the type-I magic square in the (1,1) element. Because the requirement of the type-I magic square is only the relation of $S_1= S_2 = S_3 = S_4 = S_5 =S_6 = T = T'$, these two matrices are same level of type-I magic square; however, the standard deviations of the matrices in Eq.(\ref{Eq:type-I-ex1}) and Eq.(\ref{Eq:type-I-ex2}) are $0.484$ and $0.0484$, respectively. 

We define the {\it magic index} for the type-I magic square as follows
\begin{eqnarray}
s_{\rm I} = \sqrt{\frac{\sum_{i=1}^6 (\frac{S_i}{\bar{S}}-1)^2+(\frac{T}{\bar{S}}-1)^2+(\frac{T'}{\bar{S}}-1)^2}{8}}.
\end{eqnarray}
The magic index $s_{\rm I}$ of the matrices in Eq.(\ref{Eq:type-I-ex1}) and Eq.(\ref{Eq:type-I-ex2}) are same as $0.0315$. 

The small magic index of a matrix means the matrix is much compatible with the magic square.

\

{\bf Type-II:}
If we relax the requirement of the exact magic square in Eq.(\ref{Eq:const_type-I}), we can define the quasi-magic squares. Omitting the trace $T$ in the exact magic requirement in Eq.(\ref{Eq:const_type-I}), we have a new requirement 
\begin{eqnarray}
S_1= S_2 = S_3 = S_4 = S_5 =S_6 = T',
\label{Eq:requirement-type-II}
\end{eqnarray}
for a matrix. We call a matrix with the requirement of Eq.(\ref{Eq:requirement-type-II}) the type-II magic square. For the type-II magic square, we define the following magic index 
\begin{eqnarray}
s_{\rm II} = \sqrt{\frac{\sum_{i=1}^6 (\frac{S_i}{\bar{S}}-1)^2+(\frac{T'}{\bar{S}}-1)^2}{7}},
\end{eqnarray}
where 
\begin{eqnarray}
\bar{S} = (S_1+ S_2 + S_3 + S_4 + S_5 +S_6 + T')/7.
\end{eqnarray}

\

{\bf Type-III:}
Omitting the sum $T'$ in the exact magic constraint in Eq.(\ref{Eq:const_type-I}),  we have other new requirement 
\begin{eqnarray}
S_1= S_2 = S_3 = S_4 = S_5 =S_6 = T,
\label{Eq:requirement-type-III}
\end{eqnarray}
for a matrix. We call a matrix with the requirement of Eq.(\ref{Eq:requirement-type-III}) the type-III magic square and we define the following type-III magic index
\begin{eqnarray}
s_{\rm III} = \sqrt{\frac{\sum_{i=1}^6 (\frac{S_i}{\bar{S}}-1)^2+(\frac{T}{\bar{S}}-1)^2}{7}},
\end{eqnarray}
where 
\begin{eqnarray}
\bar{S} = (S_1+ S_2 + S_3 + S_4 + S_5 +S_6 + T)/7.
\end{eqnarray}

\

{\bf Type-IV:}
We define the type-IV magic square by omitting the sum $T'$ and trace $T$ in the exact magic requirement in Eq.(\ref{Eq:const_type-I}). In this case, the requirement becomes 
\begin{eqnarray}
S_1= S_2 = S_3 = S_4 = S_5 =S_6,
\end{eqnarray}
and we define the following type-IV magic index
\begin{eqnarray}
s_{\rm IV} = \sqrt{\frac{\sum_{i=1}^6 (\frac{S_i}{\bar{S}}-1)^2}{6}},
\end{eqnarray}
where 
\begin{eqnarray}
\bar{S} = (S_1+ S_2 + S_3 + S_4 + S_5 +S_6)/6.
\end{eqnarray}

We note that  the type-I,-II,-III and -IV magic squares for the Dirac neutrinos are naive extensions of the magic texture of the Majorana neutrino mass matrix in Eq.(\ref{Eq:magicTexture}). Especially, the type-IV magic square is the most straight extension of the magic texture for Majorana neutrinos in Eq.(\ref{Eq:magicTexture}).

\

{\bf Type-V, -VI, -VII, -VIII, -XI and -X:}
We add more scenarios by omitting the one of the sum in \{$S_1$, $S_2$, $\cdots$, $S_6$\}. Namely, we define the type-V magic square with the following requirement (by omitting $S_1$)
\begin{eqnarray}
S_2 = S_3 = S_4 = S_5 =S_6 = T = T',
\end{eqnarray}
and the type-V magic index
\begin{eqnarray}
s_{\rm V} = \sqrt{\frac{\sum_{i=2}^6 (\frac{S_i}{\bar{S}}-1)^2+(\frac{T}{\bar{S}}-1)^2+(\frac{T'}{\bar{S}}-1)^2}{7}},
\label{Eq:magic_index_V}
\end{eqnarray}
where
\begin{eqnarray}
\bar{S} = (S_2 + S_3 + S_4 + S_5 + S_6 + T + T')/7.
\end{eqnarray}

Similarly, we define more five types of magic square with the following requirements 
\begin{itemize}
\item  $S_1 = S_3 = S_4 = S_5 =S_6 = T = T'$ : type-VI (by omitting $S_2$)
\item  $ S_1 = S_2 = S_4 = S_5 =S_6 = T = T'$ : type-VII (by omitting $S_3$)
\item  $ S_1 = S_2 = S_3 = S_5 =S_6 = T = T'$ : type-VII (by omitting $S_4$)
\item  $ S_1 = S_2 = S_3 = S_4 =S_6 = T = T'$ : type-IX (by omitting $S_5$)
\item  $S_1 = S_2 = S_3 = S_4 =S_5 = T = T'$ : type-X (by omitting $S_6$)
\end{itemize}
The magic indices $s_{\rm VI}$, $s_{\rm VII}$, $\cdots$ $s_{\rm X}$ are defined by the same way for $s_{\rm V}$ in Eq.(\ref{Eq:magic_index_V}).

\section{Magic square and Dirac neutrino mass matrix\label{section:mass_matrix}}
In this section, first, we show the brief review of the Dirac neutrino mass matrix and the experimental data of the neutrinos. Then, we estimate the compatibility of the magic square with the Dirac flavor neutrino mass matrix by numerical calculations.

\subsection{Neutrino mass matrix}
The minimal flavor neutrino mass matrix for Dirac neutrinos is obtained by \cite{Hagedron2005JHEP}
\begin{eqnarray}
M = \left( 
\begin{array}{ccc}
M_{ee} & M_{e\mu}  & M_{e\tau}  \\
M_{\mu e} & M_{\mu \mu}  & M_{\mu \tau}  \\
M_{\tau e} & M_{\tau\mu}  & M_{\tau\tau}  \\
\end{array}
\right) 
=
 \left( 
\begin{array}{ccc}
U_{e1} m_1 & U_{e 2} m_2 & U_{e 3} m_3 \\
U_{\mu 1} m_1 & U_{\mu 2} m_2 & U_{\mu 3} m_3 \\
U_{\tau 1} m_1 & U_{\tau 2} m_2 & U_{\tau 3} m_3 \\
\end{array}
\right) ,
\label{Eq:M-Dirac}
\end{eqnarray}
where $m_1,m_2$ and $m_3$ denote the neutrino mass eigenstates and 
\begin{eqnarray}
U_{e1} &=& c_{12}c_{13}, \quad U_{e 2} = s_{12}c_{13}, \quad U_{e 3} = s_{13} e^{-i\delta},  \\
U_{\mu 1} &=&- s_{12}c_{23} - c_{12}s_{23}s_{13} e^{i\delta}, \nonumber \\
U_{\mu 2} &=&  c_{12}c_{23} - s_{12}s_{23}s_{13}e^{i\delta}, \quad U_{\mu 3} = s_{23}c_{13}, \nonumber \\
U_{\tau 1} &=& s_{12}s_{23} - c_{12}c_{23}s_{13}e^{i\delta}, \nonumber \\
U_{\tau 2} &=& - c_{12}s_{23} - s_{12}c_{23}s_{13}e^{i\delta}, \quad U_{\tau 3} = c_{23}c_{13},\nonumber 
\end{eqnarray}
denote the elements of the Pontecorvo-Maki-Nakagawa-Sakata mixing matrix \cite{Pontecorvo1957,Pontecorvo1958,Maki1962PTP,PDG}. We used the abbreviations $c_{ij}=\cos\theta_{ij}$ and $s_{ij}=\sin\theta_{ij}$  ($i,j$=1,2,3). The Dirac CP phase is denoted by $\delta$. In this paper, we assume that the mass matrix of the charged leptons is diagonal and real.

A global analysis of current data shows the following the best-fit values of the squared mass differences $\Delta m_{ij}^2=m_i^2-m_j^2$ and the mixing angles for the so-called normal mass ordering (NO), $m_1<m_2<m_3$, of the neutrinos  \cite{Esteban2019JHEP}:
\begin{eqnarray} 
\frac{\Delta m^2_{21}}{10^{-5} {\rm eV}^2} &=& 7.39^{+0.21}_{-0.20} \quad (6.79\rightarrow 8.01), \nonumber \\
\frac{\Delta m^2_{31}}{10^{-3}{\rm eV}^2} &=& 2.528^{+0.029}_{-0.031}\quad (2.436 \rightarrow 2.618), \nonumber \\
\theta_{12}/^\circ &=& 33.82^{+0.78}_{-0.76} \quad (31.61 \rightarrow 36.27), \nonumber \\
\theta_{23}/^\circ &=& 48.6^{+1.0}_{-1.4} \quad (41.1 \rightarrow 51.3), \nonumber \\
\theta_{13}/^\circ &=& 8.60^{+0.13}_{-0.13}\quad (8.22 \rightarrow 8.98), \nonumber \\
\delta/^\circ &=& 221^{+39}_{-28}\quad (144 \rightarrow 357),
\label{Eq:neutrino_observation_NO}
\end{eqnarray}
where the $\pm$ denotes the $1 \sigma$ region and the parentheses denote the $3 \sigma$ region. On the other hands, for the so-called inverted mass ordering, $m_3 < m_1 \lesssim m_2$, we have
\begin{eqnarray} 
\frac{\Delta m^2_{21}}{10^{-5} {\rm eV}^2} &=& 7.39^{+0.21}_{-0.20} \quad (6.79 \rightarrow 8.01), \nonumber \\
\frac{\Delta m^2_{32}}{10^{-3}{\rm eV}^2} &=& -2.510^{+0.030}_{-0.031}\quad (-2.601 \rightarrow -2.416), \nonumber \\
\theta_{12}/^\circ &=& 33.82^{+0.78}_{-0.75} \quad (31.61 \rightarrow 36.27), \nonumber \\
\theta_{23}/^\circ &=& 48.4^{+1.0}_{-1.2} \quad (41.4 \rightarrow 51.3), \nonumber \\
\theta_{13}/^\circ &=& 8.64^{+0.12}_{-0.13}\quad (8.26 \rightarrow 9.02), \nonumber \\
\delta/^\circ &=& 282^{+23}_{-25}\quad (205 \rightarrow 348).
\label{Eq:neutrino_observation_IO}
\end{eqnarray}
Moreover, the following constraints  
\begin{eqnarray} 
\sum m_i < 0.12 - 0.69 ~{\rm eV},
\end{eqnarray}
from the cosmological observation of the cosmic microwave background radiation \cite{Planck2018, Capozzi2020PRD,Giusarma2016PRD,Vagnozzi2017PRD,Giusarma2018PRD} as well as 
\begin{eqnarray} 
|M_{ee}|<0.066 - 0.155 ~{\rm eV},
\end{eqnarray}
from the neutrino less double beta decay experiments \cite{GERDA2019Science,Capozzi2020PRD} are obtained.

\subsection{Numerical analysis}
\subsubsection{Setup}
To estimate the compatibility of the magic square with the Dirac flavor neutrino mass matrix by numerical calculations, we employ the following real matrix 
\begin{eqnarray}
\left(
\begin{matrix}
a & b & c \\
d & e & f \\
g & h & i
\end{matrix}
\right)= \left( 
\begin{array}{ccc}
|M_{ee}| & |M_{e\mu}|  & |M_{e\tau}|  \\
|M_{\mu e}| & |M_{\mu \mu}|  & |M_{\mu \tau}|  \\
|M_{\tau e}| & |M_{\tau\mu}|  & |M_{\tau\tau}|  \\
\end{array}
\right),
\label{Eq:magic_dirac}
\end{eqnarray}
instead of the complex Dirac mass matrix in Eq.(\ref{Eq:M-Dirac}).

In our numerical calculation, we require that the square mass differences $\Delta m^2_{ij}$, mixing angles $\theta_{ij}$ and the Dirac CP violating phase $\delta$ are varied within the $3 \sigma$ experimental ranges and the lightest neutrino mass is varied within $0.001 - 0.1$ eV. We also require that the constraints $|M_{ee}| < 0.155$ eV and $\sum m_i < 0.241$ eV (TT, TE, EE+LowE+lensing \cite{Planck2018,Singh2020EPL}) are satisfied. 

In the following subsubsections, we show the result from the numerical calculations for the ten types of magic squares. Because the type-I,-II,-III and -IV magic squares are naive extensions of the magic texture of the Majorana neutrino mass matrix in Eq.(\ref{Eq:magicTexture}), we separate our discussion into some parts. First, we will see the results for the type-I,-II,-III and -IV magic squares. Next, the results for type-V,-VI, $\cdots$ -X magic squares are presented. Moreover, we have a discussion with the results of the recently reported T2K and NOvA tension.

\subsubsection{Type-I,-II,-III and -IV}

\begin{table}[t]
\tbl{Minimum and maximum values of magic index for the type-I,-II,-III and -IV magic squares.}
{\begin{tabular}{cccc}
\hline
type & magic index & NO &  IO \\
\hline
I & $s_{\rm I}^{\rm min}$ & {\bf 0.132}  & {\bf 0.136} \\
& $s_{\rm I}^{\rm max}$ &0.783 & 0.462\\
\hline
II& $s_{\rm II}^{\rm min}$ & 0.102  & 0.107 \\
& $s_{\rm II}^{\rm max}$& 0.874 & 0.504\\
\hline
III& $s_{\rm III}^{\rm min}$ & 0.0729  & 0.0956  \\
& $s_{\rm III}^{\rm max}$& 0.739& 0.476\\
\hline
IV& $s_{\rm IV}^{\rm min}$& {\bf 0.0593}  & {\bf 0.0888} \\
& $s_{\rm IV}^{\rm max}$& 0.832& 0.525\\
\hline
\end{tabular}
\label{tbl:index}}
\end{table}

\begin{figure}[H]
\begin{center}
\includegraphics{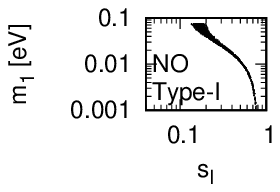}
\includegraphics{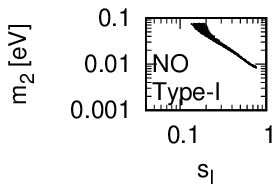}
\includegraphics{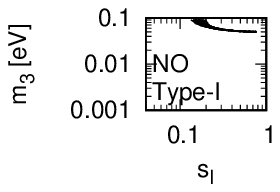}\\
\includegraphics{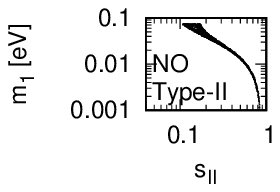}
\includegraphics{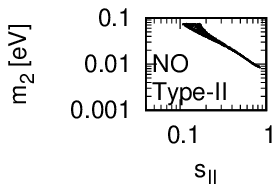}
\includegraphics{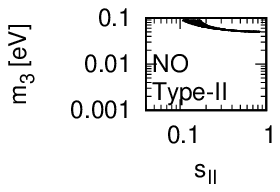}\\
\includegraphics{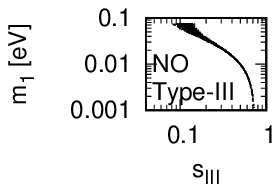}
\includegraphics{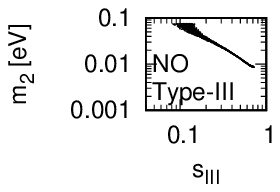}
\includegraphics{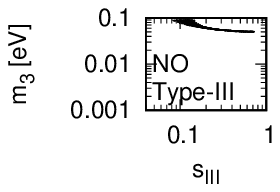}\\
\includegraphics{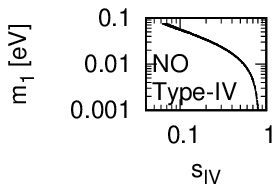}
\includegraphics{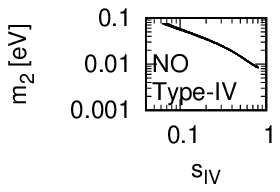}
\includegraphics{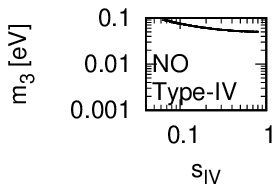}\\
\includegraphics{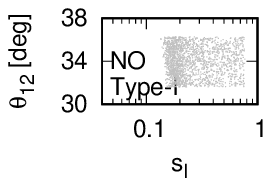}
\includegraphics{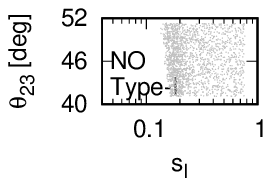}
\includegraphics{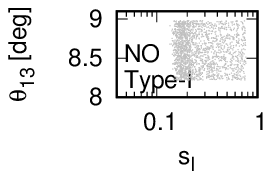}
\includegraphics{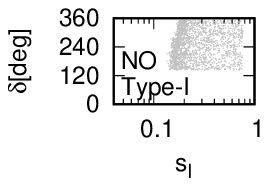} \\
\includegraphics{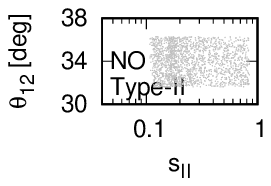}
\includegraphics{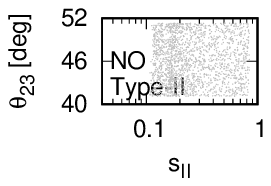}
\includegraphics{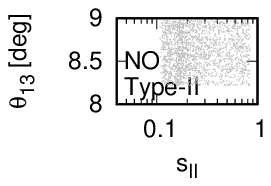}
\includegraphics{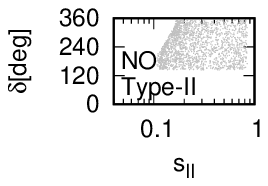} \\
\includegraphics{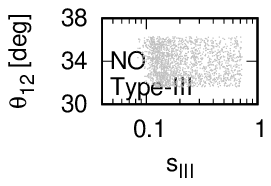}
\includegraphics{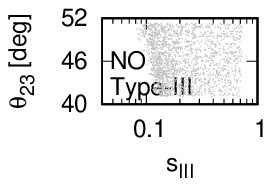}
\includegraphics{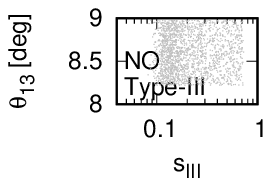}
\includegraphics{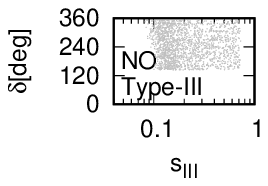} \\
\includegraphics{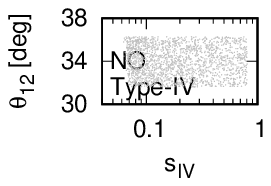}
\includegraphics{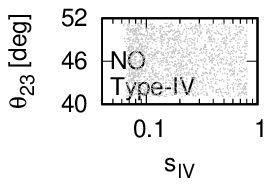}
\includegraphics{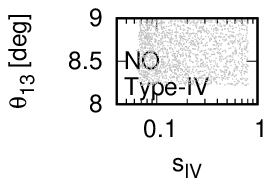}
\includegraphics{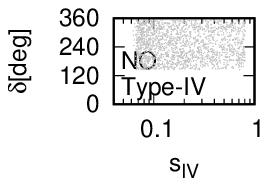}
\caption{Dependence of the neutrino parameters $m_i,\theta_{ij},\delta$ on the magic index $s$ in the case of NO for the type-I,-II,-III and -IV magic squares.}
\label{fig:NO}
\end{center}
\end{figure}

\begin{figure}[H]
\begin{center}
\includegraphics{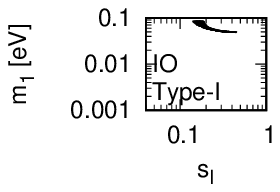}
\includegraphics{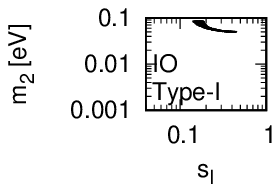}
\includegraphics{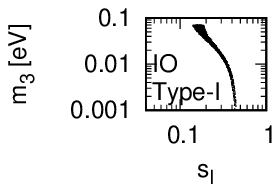}\\
\includegraphics{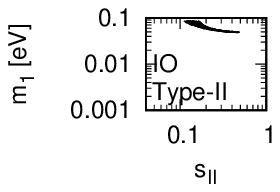}
\includegraphics{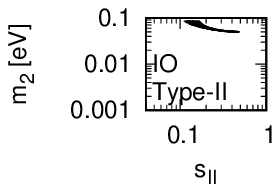}
\includegraphics{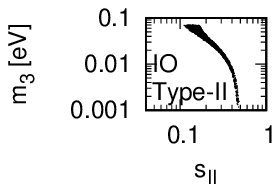}\\
\includegraphics{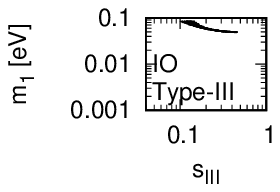}
\includegraphics{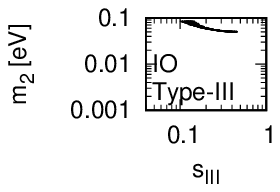}
\includegraphics{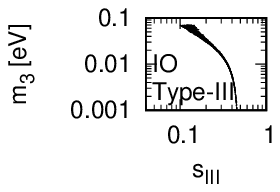}\\
\includegraphics{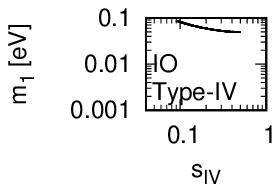}
\includegraphics{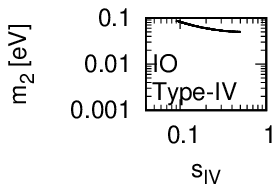}
\includegraphics{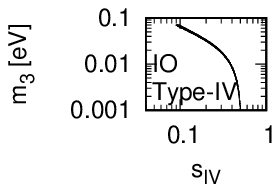}\\
\includegraphics{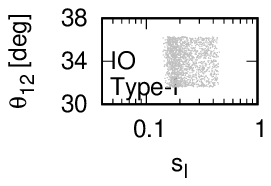}
\includegraphics{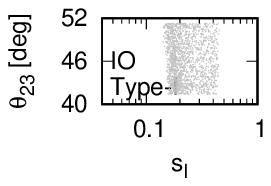}
\includegraphics{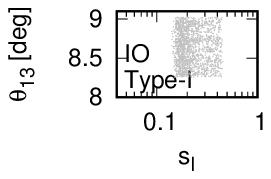}
\includegraphics{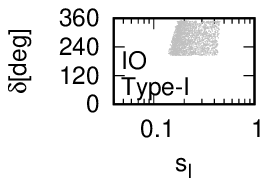} \\
\includegraphics{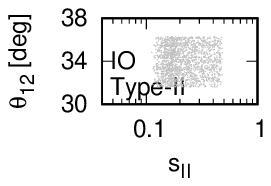}
\includegraphics{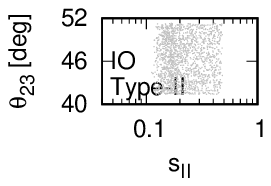}
\includegraphics{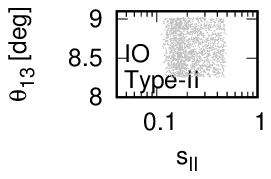}
\includegraphics{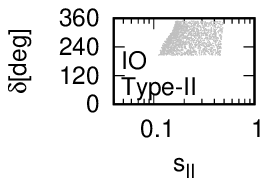} \\
\includegraphics{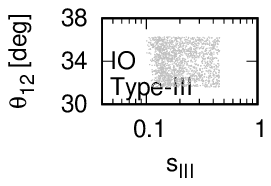}
\includegraphics{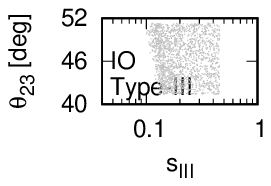}
\includegraphics{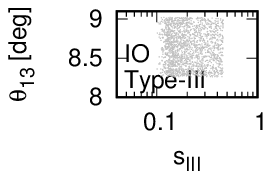}
\includegraphics{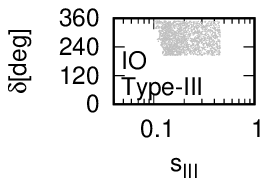} \\
\includegraphics{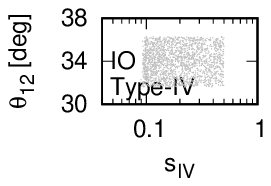}
\includegraphics{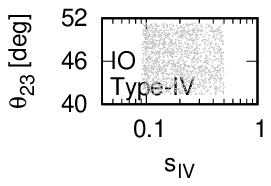}
\includegraphics{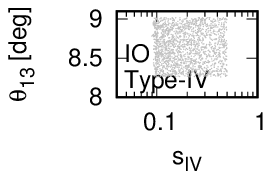}
\includegraphics{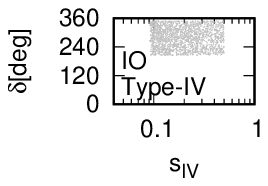}
\caption{Same as Figure \ref{fig:NO} but in the case of IO.}
\label{fig:IO}
\end{center}
\end{figure}

We show the results from the numerical calculations for the type-I,-II,-III and -IV magic squares in Tables \ref{tbl:index} - \ref{tbl:nu_param} as well as Figures \ref{fig:NO} and \ref{fig:IO}. 

Table \ref{tbl:index} shows that the minimum and maximum values of the magic indices of type-I, -II, -III and type-IV magic squares for the Dirac flavor neutrino mass matrix. We recall that the small magic index of a matrix means the matrix is much compatible with the magic square. We note the following three points:
\begin{itemize}
\item The top row in the Table \ref{tbl:index} shows that the minimum of the type-I magic index are $s_1^{\rm min}=0.132$ and $s_1^{\rm min}=0.136$ for NO and IO, respectively. Thus, the exact magic square (type-I magic square) is hardly realized for the Dirac flavor neutrino mass matrix.  
\item The smallest minimum magic index for NO(IO) is obtained in the case of type-IV as $0.0593$ ($0.0888$). Thus, if we relax the requirement of the magic squares from exact magic square (type-I) to relatively rough magic square (type-IV), some parts of the nature of the magic square are appeared for the Dirac flavor neutrino mass matrix. 
\item The magic squares prefer the normal mass ordering rather than the inverted mass ordering.
\end{itemize}
%

\begin{table}[t]
\tbl{The sums $S_1,S_2,\cdots,S_6,T,T'$ of the type-I, -II, -III and type-IV magic squares for the Dirac flavor neutrino mass matrix in the unit of ${\rm eV}$. The upper (lower) half of the table shows the sums in the case of NO (IO). For each type of magic squares, the upper (lower) row shows the sums for $s_{\rm min}$ ($s_{\rm max})$.}
{\begin{tabular}{ccccccccccc}
\hline
NO\\
\hline
type & $s$ & $S_1$ & $S_2$ & $S_3$ & $S_4$ & $S_5$ & $S_6$  & $T'$ & $T$  \\
\hline
I & $s_{\rm I}^{\rm min}$ & 0.117 & 0.133 &  0.139 &  0.120 &  0.130 &  0.138 &  {\bf 0.0971} & {\bf 0.158} \\
  & $s_{\rm I}^{\rm max}$ & 0.0128 & 0.0387 & 0.0433 & 0.00166 & 0.0144 & 0.0788 &  0.0126 & 0.0436 \\
 \hline
II & $s_{\rm II}^{\rm min}$ & 0.117 & 0.135 & 0.137 & 0.122 & 0.129 & 0.139 &  {\bf 0.100} & - \\
 & $s_{\rm II}^{\rm max}$& 0.0129 & 0.0383 & 0.0432 & 0.00168 & 0.0145 & 0.0782 &  0.0123 & - \\
 \hline
III & $s_{\rm III}^{\rm min}$ & 0.118 & 0.137 & 0.136 & 0.123 & 0.129 & 0.139 &  - & 0.149 \\
 & $s_{\rm III}^{\rm max}$& 0.0129 & 0.0387 & 0.0432 & 0.00163 & 0.0144 & 0.0788 &  - & 0.0434 \\
 \hline
IV & $s_{\rm IV}^{\rm min}$ &  0.118 & 0.137 &  0.136 &  0.123 & 0.129 & 0.139 &  - & - \\
& $s_{\rm IV}^{\rm max}$& 0.0127 & 0.0389 & 0.0432 & 0.00166 & 0.0144 & 0.0788 &  - & - \\
\hline
IO\\
\hline
type &$s$&  $S_1$ & $S_2$ & $S_3$ & $S_4$ & $S_5$ & $S_6$  & $T'$ & $T$  \\
\hline
I & $s_{\rm I}^{\rm min}$ &  0.129 &  0.126 &  0.138 &  0.137 &  0.148 &  0.107 &  {\bf 0.105} &  {\bf 0.160} \\
 & $s_{\rm I}^{\rm max}$& 0.0677 & 0.0451 & 0.0529 & 0.0798 & 0.0843 & 0.00156 &  0.0406 & 0.0619 \\
 \hline
II & $s_{\rm II}^{\rm min}$ & 0.128 & 0.132 & 0.133 & 0.138 & 0.147 & 0.109 &  {\bf 0.107} & - \\
 & $s_{\rm II}^{\rm max}$& 0.0694 & 0.0462 & 0.0543 & 0.0820 & 0.0864 & 0.00155 &  0.0414 & - \\
 \hline
III & $s_{\rm III}^{\rm min}$ & 0.128 & 0.131 & 0.134 & 0.140 & 0.146 & 0.108 &  - & 0.150 \\
 & $s_{\rm III}^{\rm max}$& 0.0683 & 0.0429 & 0.0559 & 0.0782 & 0.0873 & 0.00158 &  - & 0.0723 \\
 \hline
IV & $s_{\rm IV}^{\rm min}$ & 0.127 &  0.133 & 0.132 &  0.137&  0.147 &  0.109 &  - & - \\
 & $s_{\rm IV}^{\rm max}$& 0.0677 & 0.0430 & 0.0552 & 0.0777 & 0.0865 & 0.00163 &  - & - \\
\hline
\end{tabular}
\label{tbl:sums}}
\end{table}

Table \ref{tbl:sums} shows the sums $S_1,S_2,\cdots,S_6,T,T'$ of the type-I, -II, -III and type-IV magic squares for the Dirac flavor neutrino mass matrix in the unit of ${\rm eV}$. The upper (lower) half of the Table \ref{tbl:sums} shows the sums in the case of NO (IO). For each type of magic squares, the upper (lower) row shows the sums for $s_{\rm min}$ ($s_{\rm max})$. We note the following point:
\begin{itemize}
\item The sum of the diagonal elements $T'$ or $T$ tends to differ from other sums in the type-I and type-II. This character of $T'$ or $T$ yields the large magic index (deviation from the magic square).
\end{itemize}
%

\begin{table}[t]
\tbl{The neutrino parameters $m_i,\theta_{ij},\delta$ of the type-I, -II, -III and type-IV magic squares for the Dirac flavor neutrino mass matrix. The upper (lower) half of the table shows the sums in the case of NO (IO). For each type of magic squares, the upper (lower) row shows the sums for $s_{\rm min}$ ($s_{\rm max})$}
{\begin{tabular}{ccccccccc}
\hline
NO\\
\hline
type & $s$ & $m_1{\rm [eV]}$ & $m_2{\rm [eV]}$ & $m_3{\rm [eV]}$ & $\theta_{12}/^\circ$ &  $\theta_{23}/^\circ$  &  $\theta_{13}/^\circ$ &  $\delta/^\circ$ \\
\hline
I & $s_{\rm I}^{\rm min}$ & 0.0745  & 0.0750  & 0.0897& 36.12 & 51.19  & 8.851 & 183.9 \\ 
  & $s_{\rm I}^{\rm max}$& 0.00104 & 0.00834 & 0.0511& 33.14 & 41.39 & 8.400 & 345.7 \\
\hline
II & $s_{\rm II}^{\rm min}$& 0.0746  & 0.0750  & 0.0894& 35.56  & 41.17  & 8.964 & 181.4  \\
   & $s_{\rm II}^{\rm max}$& 0.00105& 0.00838 & 0.0507& 34.92 & 41.12 & 8.270 & 345.9 \\
\hline
III & $s_{\rm III}^{\rm min}$& 0.0748  & 0.0754  & 0.0899& 36.20  & 51.10  &  8.866  & 340.2  \\
   & $s_{\rm III}^{\rm max}$& 0.00101& 0.00832 & 0.0511& 35.43 & 41.48 & 8.277 & 336.7\\
\hline
IV & $s_{\rm IV}^{\rm min}$& 0.0751  & 0.0756  & 0.0899& 35.88  & 51.17 & 8.980  & 354.5\\
  & $s_{\rm IV}^{\rm max}$& 0.00104& 0.00832 & 0.0512& 32.88 & 41.51 & 8.294 & 325.1\\
\hline
IO\\
\hline
type  & $s$ & $m_1{\rm [eV]}$ & $m_2{\rm [eV]}$ & $m_3{\rm [eV]}$ & $\theta_{12}/^\circ$ &  $\theta_{23}/^\circ$  &  $\theta_{13}/^\circ$ &  $\delta/^\circ$ \\
\hline
I & $s_{\rm I}^{\rm min}$& 0.0852  & 0.0856 & 0.0691& 36.21  & 51.20 & 8.982  &  205.1\\
& $s_{\rm I}^{\rm max}$& 0.0487 & 0.0494& 0.00101& 35.50  & 51.08& 8.619 & 343.9 \\
\hline
II & $s_{\rm II}^{\rm min}$& 0.0852  & 0.0856  & 0.0701& 33.23 & 41.41  & 8.881 & 205.3  \\
 & $s_{\rm II}^{\rm max}$& 0.0500& 0.0506 & 0.00101& 35.66 &  51.10 & 8.822 &  347.4\\
 \hline
III & $s_{\rm III}^{\rm min}$ & 0.0851  & 0.0855  & 0.0699& 35.74  & 50.98  & 8.947  & 343.5  \\
& $s_{\rm III}^{\rm max}$& 0.0498& 0.0506& 0.00102& 31.62& 51.16& 8.594& 210.3\\
\hline
IV & $s_{\rm IV}^{\rm min}$& 0.0852  & 0.0857 & 0.0701& 32.25  & 48.50  & 8.906  & 329.1  \\
& $s_{\rm IV}^{\rm max}$& 0.0494& 0.0500  & 0.00106& 32.14& 50.68 &  8.959 & 208.5\\
\hline
\end{tabular}
\label{tbl:nu_param}}
\end{table}

Table \ref{tbl:nu_param} shows the neutrino parameters $m_i,\theta_{ij},\delta$ of the type-I, -II, -III and type-IV magic squares for the Dirac flavor neutrino mass matrix. As same as Table \ref{tbl:sums}, the upper (lower) half of the Table \ref{tbl:nu_param} shows the sums in the case of NO (IO). For each type of magic squares, the upper (lower) row shows the sums for $s_{\rm min}$ ($s_{\rm max})$. The following ranges of the neutrino parameters are roughly favored for the magic squares:
\begin{eqnarray} 
m_i/{\rm eV} &\sim &0.07 - 0.09, \nonumber  \\
\theta_{12}/^\circ &\sim& 36, \nonumber  \\
\theta_{23}/^\circ &\sim& 51, {\rm (for ~ type-I,-III,-IV)}, \quad \sim 41, {\rm (for ~ type-II)}, \nonumber  \\
\theta_{13}/^\circ &\sim& 8.9 \nonumber  \\
\delta/^\circ &\sim& 181-184, {\rm (for ~ type-I,-II)}, \quad \sim 340, {\rm (for ~ type-III, IV)}, 
\end{eqnarray}
for NO and
\begin{eqnarray} 
m_i/{\rm eV} &\sim & 0.07 - 0.09, \nonumber  \\
\theta_{12}/^\circ &\sim& 33 - 36, \nonumber  \\
\theta_{23}/^\circ &\sim& 51, {\rm (for ~ type-I,-III,-IV)}, \quad \sim 41, {\rm (for ~ type-II)}, \nonumber  \\
\theta_{13}/^\circ &\sim& 8.9 \nonumber  \\
\delta/^\circ &\sim& 205, {\rm (for ~ type-I,-II)}, \quad \sim 329 - 344, {\rm (for ~ type-III, IV)}, 
\end{eqnarray}
for IO.

Figure \ref{fig:NO} and Figure \ref{fig:IO} show that the dependence of the neutrino parameters $m_i,\theta_{ij},\delta$ on the magic index $s$ in the case of NO and IO, respectively. We note the following four points:
\begin{itemize}
\item The large neutrino masses yield small magic indices (the large masses are favorable for magic squares).
\item The minimum of $s_{\rm I}$, $s_{\rm II}$ and $s_{\rm III}$ for type-I, -II and -III are obtained with large $\theta_{12}$, $\theta_{23}$ and $\theta_{13}$.
\item The minimum of $s_{\rm I}$ and $s_{\rm II}$ for type-I and  type-II are obtained with small $\delta$.
\item The minimum of $s_{\rm III}$ for type-III is obtained with large $\delta$.
\end{itemize}
for both NO and IO.

We would like to emphasis again the following remarkable result of our numerical calculations:
\begin{itemize}
\item All types of the magic squares prefer the normal mass ordering rather than the inverted mass ordering for the type-V,-VI, $\cdots$, -X magic squares (see Table \ref{tbl:index}).
\end{itemize}
Although the neutrino mass ordering (either NO or IO) is not determined experimentaly, a global analysis shows that the preference for the normal mass ordering is mostly due to neutrino oscillation measurements \cite{Salas2018PLB, Capozzi2020PRD,Aartsen2020PRD}. The theoretical origin of the mass ordering of neutrinos is still big problem. There is a possibility that the origin of the normal mass ordering is the magic nature of the neutrinos. This conclusion becomes more rigid in the next subsubsection.

\clearpage

\subsubsection{Type-V,-VI, $\cdots$, -X}

\begin{figure}[H]
\begin{center}
\includegraphics{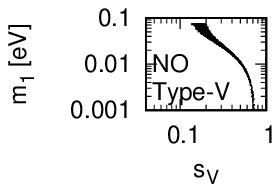}
\includegraphics{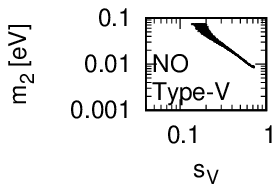}
\includegraphics{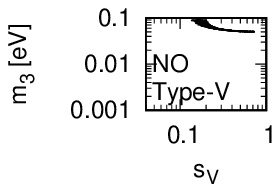}\\
\includegraphics{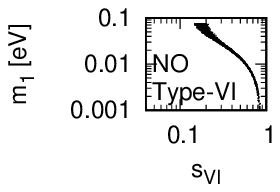}
\includegraphics{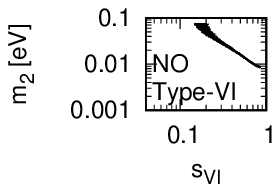}
\includegraphics{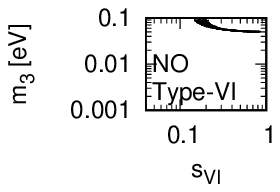}\\
\includegraphics{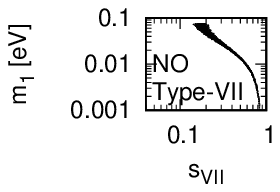}
\includegraphics{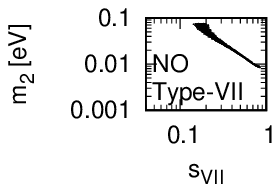}
\includegraphics{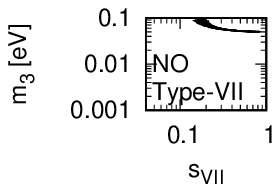}\\
\includegraphics{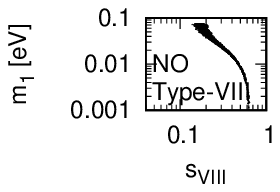}
\includegraphics{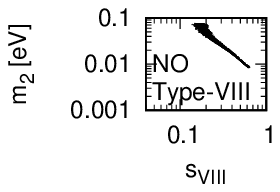}
\includegraphics{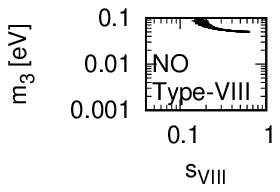}\\
\includegraphics{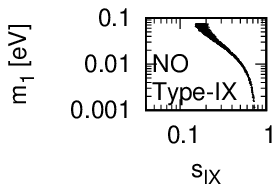}
\includegraphics{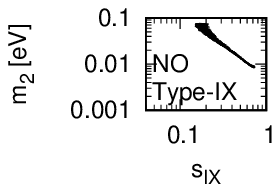}
\includegraphics{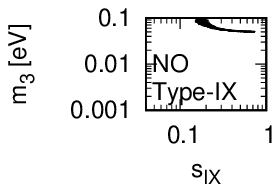}\\
\includegraphics{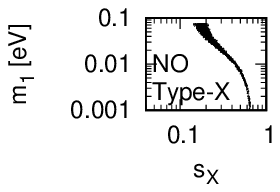}
\includegraphics{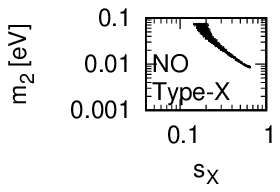}
\includegraphics{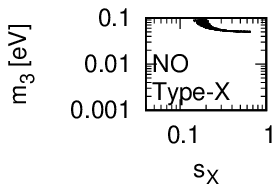}\\
\caption{Dependence of the neutrino mass $m_i$ on the magic index $s$ in the case of NO for the type-V,-VI, $\cdots$, -X magic squares.}
\label{fig:NO-mass-V-X}
\end{center}
\end{figure}

We show the results from the numerical calculations for the type-V,-VI, $\cdots$, -X magic squares in Tables \ref{tbl:index_for_V_X} - \ref{tbl:nu_param_for_V_X} as well as Figures \ref{fig:NO-mass-V-X} and \ref{fig:IO-mixing-V-X}. 

Table \ref{tbl:index_for_V_X} shows that the minimum and maximum values of the magic indices of type-V,-VI, $\cdots$, -X magic squares for the Dirac flavor neutrino mass matrix. Again, we recall that the small magic index of a matrix means the matrix is much compatible with the magic square. We note the following two points:
\begin{itemize}
\item The smallest minimum magic index for NO is obtained in the case of type-V as $0.133$ and for IO is obtained in the case of type-X as $0.120$. These minimum magic indices are lager than the type-IV magic indices $0.0593$ (NO) and $0.0888$ (IO) in Table \ref{tbl:index}. Thus, the most compatible type of magic square for Dirac neutrinos is the type-IV which is the most straight extension of the Majorana magic mass matrix texture.
\item The magic squares prefer the normal mass ordering rather than the inverted mass ordering except with type-IX and type-X. According to the results shown in Tables \ref{tbl:index} and \ref{tbl:index_for_V_X}, we conclude that the possibility that the origin of the normal mass ordering is the magic nature of the neutrinos becomes more rigid.
\end{itemize}
%

\begin{table}[t]
\tbl{Minimum and maximum values of magic index for the type-V,-VI, $\cdots$, -X magic squares.}
{\begin{tabular}{cccc}
\hline
type & magic index & NO &  IO \\
\hline
V & $s_{\rm V}^{\rm min}$ & {\bf 0.133} & 0.144 \\
& $s_{\rm V}^{\rm max}$ & 0.742 & 0.492 \\
\hline
VI & $s_{\rm VI}^{\rm min}$ & 0.140  & 0.143 \\
& $s_{\rm VI}^{\rm max}$ & 0.873 & 0.480 \\
\hline
VII & $s_{\rm VII}^{\rm min}$ & 0.138  & 0.144 \\
& $s_{\rm VII}^{\rm max}$ & 0.864 & 0.483 \\
\hline
VIII & $s_{\rm VIII}^{\rm min}$ & 0.137  & 0.145 \\
& $s_{\rm VIII}^{\rm max}$ & 0.655 & 0.482 \\
\hline
IX & $s_{\rm IX}^{\rm min}$ & 0.142  & 0.137 \\
& $s_{\rm IX}^{\rm max}$ & 0.752 & 0.468 \\
\hline
X & $s_{\rm X}^{\rm min}$ & 0.139  & {\bf 0.120} \\
& $s_{\rm X}^{\rm max}$ & 0.674 & 0.250 \\
\hline
\end{tabular}
\label{tbl:index_for_V_X}}
\end{table}

Table \ref{tbl:sums_for_V_X} shows the sums $S_1,S_2,\cdots,S_6,T,T'$ of the type-V,-VI, $\cdots$, -X magic squares for the Dirac flavor neutrino mass matrix in the unit of ${\rm eV}$. The upper (lower) half of the Table \ref{tbl:sums_for_V_X} shows the sums in the case of NO (IO). For each type of magic squares, the upper (lower) row shows the sums for $s_{\rm min}$ ($s_{\rm max})$. We note the following point:
\begin{itemize}
\item The magnitudes of the each entries in Table \ref{tbl:sums_for_V_X} for the type-V,-VI, $\cdots$, -X magic squares are almost same with the magnitudes of the each entries in Table \ref{tbl:sums} for the type-I,-II, $\cdots$, -IV magic squares.
\end{itemize}

\begin{table}[t]
\tbl{The sums $S_1,S_2,\cdots,S_6,T,T'$ of the type-V,-VI, $\cdots$, -X magic squares for the Dirac flavor neutrino mass matrix in the unit of ${\rm eV}$. The upper (lower) half of the table shows the sums in the case of NO (IO). For each type of magic squares, the upper (lower) row shows the sums for $s_{\rm min}$ ($s_{\rm max})$.}
{\begin{tabular}{ccccccccccc}
\hline
NO\\
\hline
type & $s$ & $S_1$ & $S_2$ & $S_3$ & $S_4$ & $S_5$ & $S_6$  & $T'$ & $T$  \\
\hline
V & $s_{\rm V}^{\rm min}$ & - & 0.133 & 0.138 &  0.120 & 0.130 & 0.139 &  0.0970 & 0.158  \\
  & $s_{\rm V}^{\rm max}$ & - & 0.0437 & 0.0379 &  0.001657 & 0.0147 & 0.0783 & 0.0118  & 0.0364 \\
 \hline
VI & $s_{\rm VI}^{\rm min}$ & 0.117 & - & 0.139 &  0.120 & 0.130 & 0.138 &  0.0975 & 0.159 \\
  & $s_{\rm VI}^{\rm max}$ & 0.0127 & - & 0.0382 &  0.00172 & 0.0143 & 0.0781 &  0.0116 & 0.0369 \\
 \hline
VII & $s_{\rm VII}^{\rm min}$ & 0.118 & 0.133 & - &  0.121 & 0.131 & 0.139 &  0.0980 & 0.159 \\
  & $s_{\rm VII}^{\rm max}$ & 0.0129 & 0.0431 & - & 0.00163 & 0.0142 & 0.0787 & 0.0119  & 0.0371 \\
 \hline
VIII & $s_{\rm VIII}^{\rm min}$ &0.117 & 0.133 & 0.139 &  -& 0.130 & 0.139 & 0.0973  & 0.159 \\
  & $s_{\rm VIII}^{\rm max}$ & 0.0129 &0.0431 & 0.0374 &  - & 0.0143 & 0.0775 & 0.0116  & 0.0358 \\
\hline
IX & $s_{\rm IX}^{\rm min}$ & 0.115 & 0.131 & 0.137 &  0.118 & - & 0.137 &  0.0954 & 0.156 \\
  & $s_{\rm IX}^{\rm max}$ & 0.0131 & 0.0435 & 0.0378 &  0.00165 & - & 0.0781 & 0.0118  & 0.0363 \\
\hline
X & $s_{\rm X}^{\rm min}$ & 0.0117 & 0.133 & 0.138 & 0.119 & 0.123 & - &  0.0972 & 0.159 \\
  & $s_{\rm X}^{\rm max}$ & 0.0126 & 0.0387 & 0.0429 & 0.00163 & 0.0146 & - & 0.0125  & 0.0431 \\
\hline
IO\\
\hline
type &$s$&  $S_1$ & $S_2$ & $S_3$ & $S_4$ & $S_5$ & $S_6$  & $T'$ & $T$  \\
\hline
V & $s_{\rm V}^{\rm min}$ & - & 0.126 & 0.136 &  0.137 & 0.147 & 0.106 & 0.103  & 0.159 \\
  & $s_{\rm V}^{\rm max}$ & - & 0.0453 & 0.0530 &  0.0799 & 0.0847 & 0.00156 & 0.0407  & 0.0627 \\
 \hline
VI & $s_{\rm VI}^{\rm min}$ & 0.128 & - & 0.137 &  0.137 & 0.147 & 0.106 &  0.104 & 0.159 \\
  & $s_{\rm VI}^{\rm max}$ & 0.0677 & - & 0.0449 &  0.0779 & 0.0865& 0.00159 & 0.0421  & 0.0713 \\
 \hline
VII & $s_{\rm VII}^{\rm min}$ & 0.129 & 0.127 & - &  0.138 & 0.148 & 0.108 & 0.105  & 0.161 \\
  & $s_{\rm VII}^{\rm max}$ & 0.0684 & 0.0456 & - &  0.0806 & 0.0854 & 0.00168 & 0.0409  & 0.0631 \\
 \hline
VIII & $s_{\rm VIII}^{\rm min}$ &0.128 & 0.126 & 0.137 &  - & 0.147 & 0.107 & 0.103  & 0.159 \\
  & $s_{\rm VIII}^{\rm max}$ & 0.0668 & 0.0528 & 0.0440 &  - & 0.0853 & 0.00166 & 0.0420  & 0.0709 \\
\hline
IX & $s_{\rm IX}^{\rm min}$ & 0.128 & 0.126 & 0.137 & 0.137  & - & 0.107 & 0.104 & 0.160 \\
  & $s_{\rm IX}^{\rm max}$ & 00.0678 & 0.0449 & 0.0532 &  0.0801 & - & 0.00154 & 0.0408  & 0.0612 \\
\hline
X & $s_{\rm X}^{\rm min}$ & 0.121 & 0.119 & 0.129 &  0.130 & 0.140 & - &  0.0989 & 0.151 \\
  & $s_{\rm X}^{\rm max}$ & 0.0681 & 0.0458 & 0.0531 &  0.0802 & 0.0850 & - & 0.0407  & 0.0629 \\
\hline
\end{tabular}
\label{tbl:sums_for_V_X}}
\end{table}


\begin{table}[t]
\tbl{The neutrino parameters $m_i,\theta_{ij},\delta$ of the type-V,-VI, $\cdots$, -X magic squares for the Dirac flavor neutrino mass matrix. The upper (lower) half of the table shows the sums in the case of NO (IO). For each type of magic squares, the upper (lower) row shows the sums for $s_{\rm min}$ ($s_{\rm max})$}
{\begin{tabular}{ccccccccc}
\hline
NO\\
\hline
type & $s$ & $m_1{\rm [eV]}$ & $m_2{\rm [eV]}$ & $m_3{\rm [eV]}$ & $\theta_{12}/^\circ$ &  $\theta_{23}/^\circ$  &  $\theta_{13}/^\circ$ &  $\delta/^\circ$ \\
\hline
V & $s_{\rm V}^{\rm min}$ & 0.0743  & 0.0748  & 0.0898 & 36.25 & 51.24  & 8.914 & 170.7 \\ 
  & $s_{\rm V}^{\rm max}$& 0.00102 & 0.00858 & 0.0509 & 34.18 & 51.19 & 8.377 & 330.6 \\
\hline
VI & $s_{\rm VI}^{\rm min}$ & 0.0746  & 0.0752  & 0.0895 & 35.97 & 51.23  & 8.972 & 180.6 \\ 
  & $s_{\rm VI}^{\rm max}$& 0.00107 & 0.00834 & 0.0508 & 33.10 & 50.47 & 8.277 & 340.5 \\
\hline
VII & $s_{\rm VII}^{\rm min}$ & 0.0750  & 0.0755  & 0.0899 & 36.08 & 51.17  & 8.970 & 172.8 \\ 
  & $s_{\rm VII}^{\rm max}$& 0.00101 & 0.00830 & 0.0510 & 33.27 & 50.16 & 8.576 & 350.4 \\
\hline
VIII & $s_{\rm VIII}^{\rm min}$ & 0.0748  & 0.0753  & 0.0897 & 36.20 & 50.77  & 8.872 & 164.8 \\ 
  & $s_{\rm VIII}^{\rm max}$& 0.00102 & 0.0839 & 0.0504 & 34.27 & 51.25 & 8.457 & 351.8 \\
\hline
IX & $s_{\rm IX}^{\rm min}$ & 0.0734  & 0.0739  & 0.0889 & 36.23 & 51.12  & 8.711 & 183.0 \\ 
  & $s_{\rm IX}^{\rm max}$& 0.00101 & 0.00859 & 0.0508 & 34.63 & 51.10 & 8.438 & 335.8 \\
\hline
X & $s_{\rm X}^{\rm min}$ & 0.0743  & 0.0748  & 0.0889 & 36.05 & 51.05  & 8.959 & 182.9 \\ 
  & $s_{\rm X}^{\rm max}$& 0.00104 & 0.00844 & 0.0506 & 31.78 & 41.71 & 8.361 & 356.4 \\
\hline
IO\\
\hline
type  & $s$ & $m_1{\rm [eV]}$ & $m_2{\rm [eV]}$ & $m_3{\rm [eV]}$ & $\theta_{12}/^\circ$ &  $\theta_{23}/^\circ$  &  $\theta_{13}/^\circ$ &  $\delta/^\circ$ \\
\hline
V & $s_{\rm V}^{\rm min}$ & 0.0843  & 0.0847  & 0.0689 & 36.07 & 50.78  & 8.825 & 208.8 \\ 
  & $s_{\rm V}^{\rm max}$& 0.0489 & 0.0496 & 0.00101 & 34.82 & 51.06 & 8.795 & 341.3 \\
\hline
VI & $s_{\rm VI}^{\rm min}$ & 0.0847  & 0.0851  & 0.0689 & 36.10 & 51.06  & 8,865 & 205.5 \\ 
  & $s_{\rm VI}^{\rm max}$& 0.0494 & 0.0501 & 0.00103 & 31.80 & 41.88 & 8.847 & 343.6 \\
\hline
VII & $s_{\rm VII}^{\rm min}$ & 0.0853  & 0.0857  & 0.0696 & 35.97 & 50.76  & 9.009 & 206.9 \\ 
  & $s_{\rm VII}^{\rm max}$& 0.0493 & 0.05000 & 0.00109 & 34.86 & 51.29 & 8.918 & 342.4 \\
\hline
VIII & $s_{\rm VIII}^{\rm min}$ & 0.0846 & 0.0851  & 0.0689 & 35.92 & 50.93  & 9.015 & 217.0 \\ 
  & $s_{\rm VIII}^{\rm max}$& 0.0487 & 0.0494 & 0.00107 & 31.68 & 41.45 & 8.480 & 340.9\\
\hline
IX & $s_{\rm IX}^{\rm min}$ & 0.0848  & 0.0853  & 0.0694 & 35.91 & 51.11  & 8.835 & 206.7 \\ 
  & $s_{\rm IX}^{\rm max}$& 0.0486 & 0.0494 & 0.00100 & 36.14 & 51.18 & 8.435 & 347.2 \\
\hline
X & $s_{\rm X}^{\rm min}$ & 0.0805  & 0.0810  & 0.0643 & 36.19 & 50.89  & 8.895 & 207.8 \\ 
  & $s_{\rm X}^{\rm max}$& 0.0491 & 0.0498 & 0.00120 & 35.04 & 50.84 & 8.858 & 346.7 \\
\hline
\end{tabular}
\label{tbl:nu_param_for_V_X}}
\end{table}

Table \ref{tbl:nu_param_for_V_X} shows the neutrino parameters $m_i,\theta_{ij},\delta$ of the type-V,-VI, $\cdots$, -X magic squares for the Dirac flavor neutrino mass matrix. As same as Table \ref{tbl:sums_for_V_X}, the upper (lower) half of the Table \ref{tbl:nu_param_for_V_X} shows the sums in the case of NO (IO). For each type of magic squares, the upper (lower) row shows the sums for $s_{\rm min}$ ($s_{\rm max})$. The following ranges of the neutrino parameters are roughly favored for the magic squares:
\begin{eqnarray} 
m_i/{\rm eV} &\sim &0.07 - 0.09, \nonumber  \\
\theta_{12}/^\circ &\sim& 36, \nonumber  \\
\theta_{23}/^\circ &\sim& 51, \nonumber  \\
\theta_{13}/^\circ &\sim& 8.9, \nonumber  \\
\delta/^\circ &\sim& 165 -180, 
\end{eqnarray}
for NO and
\begin{eqnarray} 
m_i/{\rm eV} &\sim & 0.07 - 0.085, \nonumber  \\
\theta_{12}/^\circ &\sim& 36, \nonumber  \\
\theta_{23}/^\circ &\sim& 51, \nonumber  \\
\theta_{13}/^\circ &\sim& 8.8 - 9.0 \nonumber  \\
\delta/^\circ &\sim& 205 - 217, 
\end{eqnarray}
for IO. In comparison with the type-I, -II $\cdots$ -IV magic squares, the variance of each neutrino parameters is small in the case of type-V,-VI, $\cdots$, -X magic squares.

\begin{figure}[H]
\begin{center}
\includegraphics{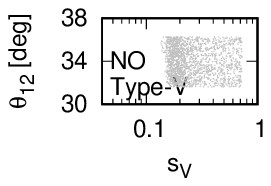}
\includegraphics{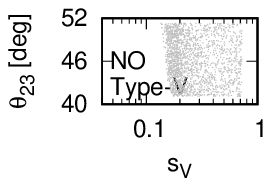}
\includegraphics{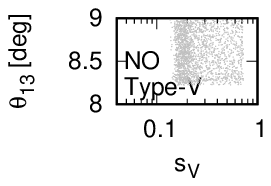}
\includegraphics{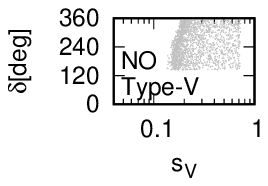} \\
\includegraphics{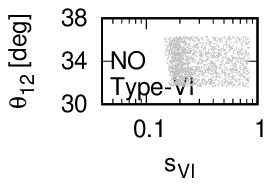}
\includegraphics{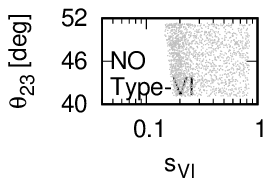}
\includegraphics{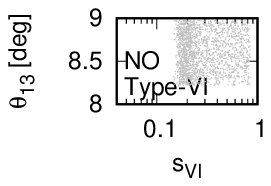}
\includegraphics{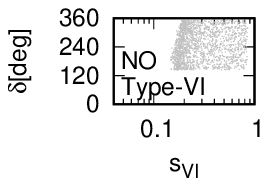} \\
\includegraphics{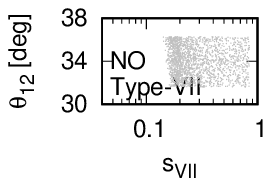}
\includegraphics{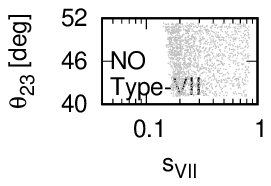}
\includegraphics{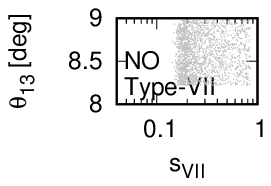}
\includegraphics{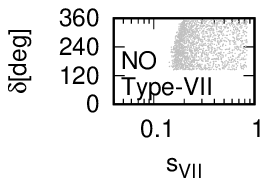} \\
\includegraphics{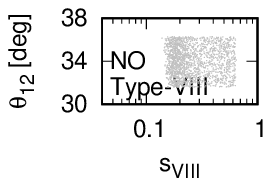}
\includegraphics{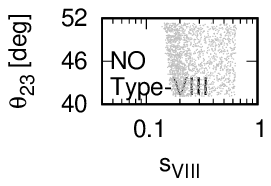}
\includegraphics{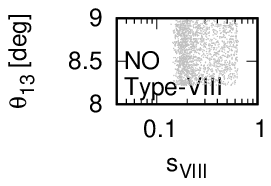}
\includegraphics{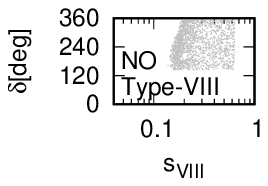} \\
\includegraphics{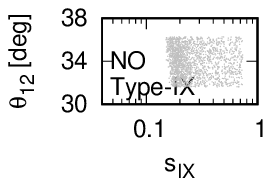}
\includegraphics{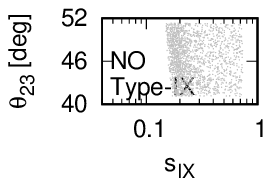}
\includegraphics{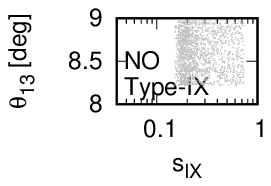}
\includegraphics{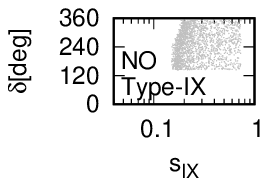} \\
\includegraphics{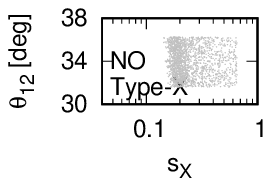}
\includegraphics{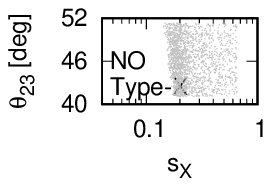}
\includegraphics{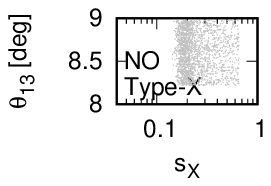}
\includegraphics{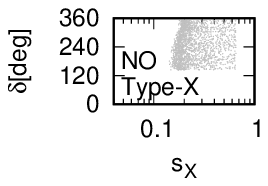}
\caption{Dependence of the neutrino mixing angles and phase $\theta_{ij},\delta$ on the magic index $s$ in the case of NO for the the type-V,-VI, $\cdots$, -X magic squares.}
\label{fig:NO-mixing-V-X}
\end{center}
\end{figure}

\begin{figure}[H]
\begin{center}
\includegraphics{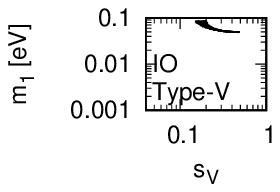}
\includegraphics{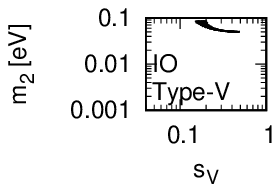}
\includegraphics{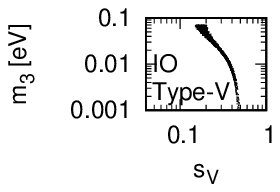}\\
\includegraphics{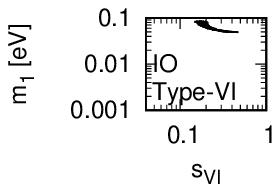}
\includegraphics{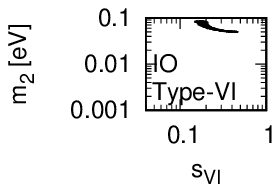}
\includegraphics{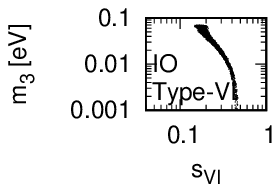}\\
\includegraphics{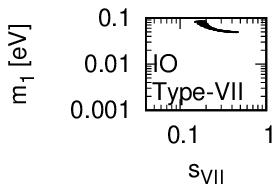}
\includegraphics{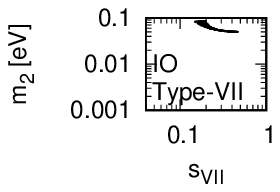}
\includegraphics{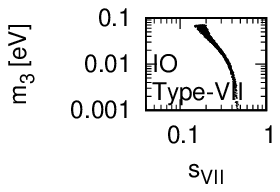}\\
\includegraphics{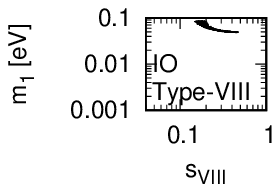}
\includegraphics{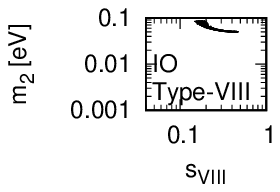}
\includegraphics{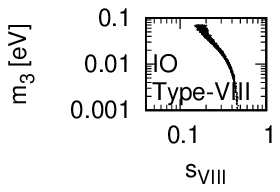}\\
\includegraphics{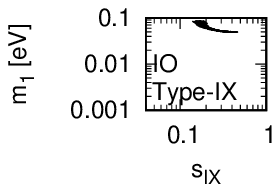}
\includegraphics{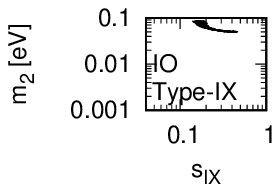}
\includegraphics{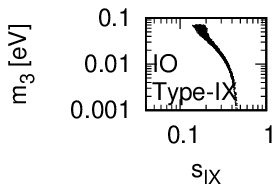}\\
\includegraphics{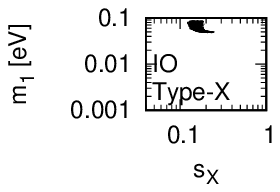}
\includegraphics{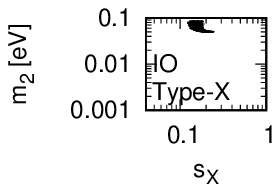}
\includegraphics{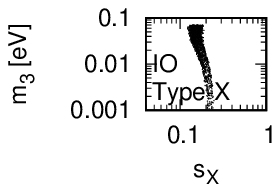}\\
\caption{Same as Figure \ref{fig:NO-mass-V-X} but in the case of IO.}
\label{fig:IO-mass-V-X}
\end{center}
\end{figure}

\begin{figure}[H]
\begin{center}
\includegraphics{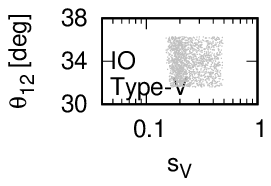}
\includegraphics{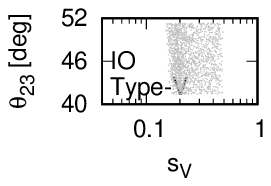}
\includegraphics{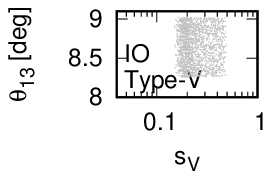}
\includegraphics{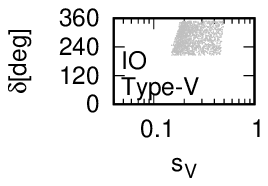} \\
\includegraphics{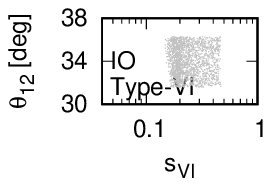}
\includegraphics{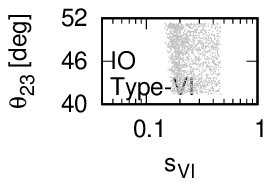}
\includegraphics{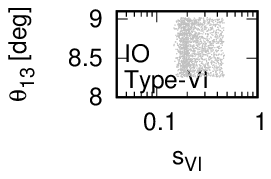}
\includegraphics{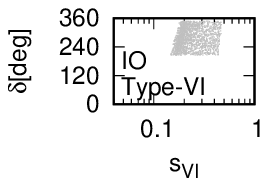} \\
\includegraphics{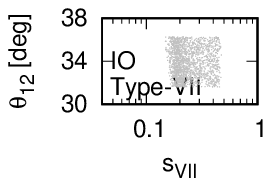}
\includegraphics{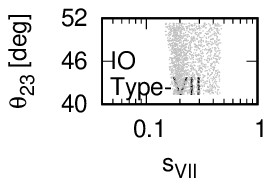}
\includegraphics{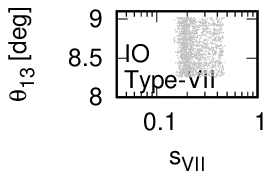}
\includegraphics{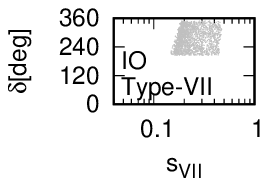} \\
\includegraphics{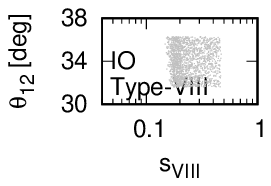}
\includegraphics{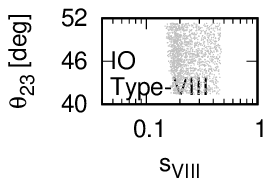}
\includegraphics{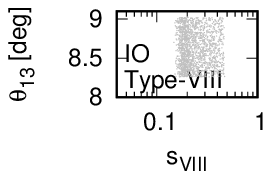}
\includegraphics{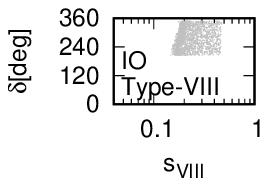} \\
\includegraphics{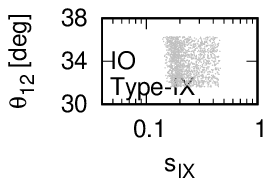}
\includegraphics{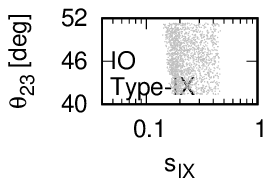}
\includegraphics{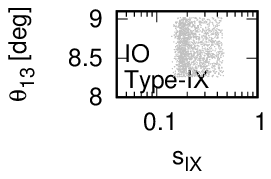}
\includegraphics{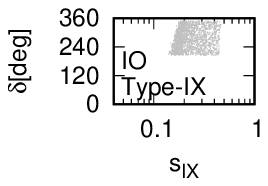} \\
\includegraphics{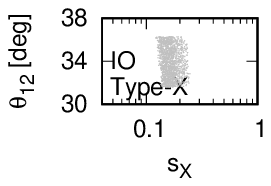}
\includegraphics{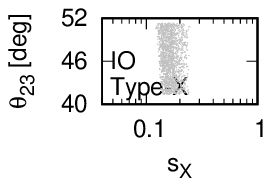}
\includegraphics{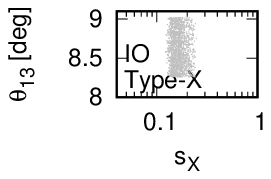}
\includegraphics{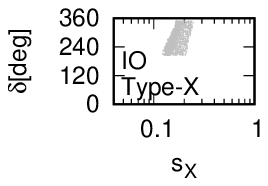}
\caption{Same as Figure \ref{fig:NO-mixing-V-X} but in the case of IO.}
\label{fig:IO-mixing-V-X}
\end{center}
\end{figure}

Figure \ref{fig:NO-mass-V-X} and Figure \ref{fig:NO-mixing-V-X} show that the dependence of the neutrino parameters $m_i,\theta_{ij},\delta$ on the magic index $s$ in the case of NO. Figure \ref{fig:IO-mass-V-X} and Figure \ref{fig:IO-mixing-V-X} show the same as Figure \ref{fig:NO-mass-V-X} and Figure \ref{fig:NO-mixing-V-X} but in the case of IO. We note the following two points:
\begin{itemize}
\item The large neutrino masses yield small magic indices (the large masses are favorable for magic squares).
\item The minimum of magic index $s_i$ $(i={\rm V,VI,\cdots, X})$ is obtained with large $\theta_{12}$, $\theta_{23}$ and $\theta_{13}$ as well as with small $\delta$.
\end{itemize}
for both NO and IO.

\subsubsection{T2K and NOvA tension}
Recently, T2K and NOvA reported a tension in the measurement of $\delta$ and $\sin^2\theta_{23}$ for the normal mass ordering of neutrinos \cite{Himmel2020,Dunne2020}. Both experiments favor the upper octant of $\theta_{23}$; however, the T2K data shows $\delta >\pi$ against the NOvA result $\delta < \pi$. Now, we have a discussion with the results of the latest T2K and NOvA experiments.

Table \ref{tbl:T2K_NOvA} shows the mixing angle $\theta_{23}$ and the Dirac CP phase $\delta$ for minimum magic index of the type-I, -II, $\cdots$, -X magic squares for the Dirac flavor neutrino mass matrix with NO. The symbol ``O" (``$\times$") means the corresponding case is favorable (unfavorable) with recent T2K or NOvA data.

Since the most compatible type of magic square is the type-IV (which has the smallest minimum magic index $s_{\rm IV}^{\rm min} = 0.0593$), it seems that the T2K result is more favorable than the NOvA result in the context of the magic square.

\begin{table}[t]
\tbl{The mixing angle $\theta_{23}$ and the Dirac CP phase $\delta$ for minimum magic index of the type-I, -II, $\cdots$, -X magic squares for the Dirac flavor neutrino mass matrix with NO. The symbol ``O" (``$\times$") means the corresponding case is favorable (unfavorable) with recent T2K or NOvA data.}
{\begin{tabular}{ccccccccccc}
\hline
 & I & II &III &{\bf IV} & V & VI  & VII & VIII & IX & X\\
\hline
$s_i^{\rm min}$ & 0.132 & 0.102  & 0.0729 & {\bf 0.0593} & 0.133 & 0.140 & 0.138 & 0.137 & 0.142 &0.139 \\ 
\hline
$\theta_{23}/^\circ$ & 51.19 & 41.17  & 51.10 & {\bf 51.17} & 51.24  & 51.23 & 51.17 & 50.77 & 51.12 &51.05 \\ 
$\delta/^\circ$ & 183.9 & 181.4  & 340.2 & {\bf 354.5} & 170.7  & 180.6 & 172.8 & 164.8 & 183.0 &182.9 \\
T2K & O & $\times$  & O & {\bf O} & $\times$  & O & $\times$ & $\times$ & O &O \\
NOvA & $\times$ & $\times$  & $\times$ & $\boldmath{\times}$ & O  & $\times$ & O & O & $\times$ &$\times$ \\
\hline
\end{tabular}
\label{tbl:T2K_NOvA}}
\end{table}

\clearpage

\clearpage

\section{Summary\label{section:summary}}
The magic texture is one of the successful textures of the flavor neutrino mass matrix for the Majorana type neutrinos. The name ``magic" is inspired by the nature of the magic square. In this paper, we have estimated the compatibility of the magic square with the Dirac, instead of the Majorana, flavor neutrino mass matrix by numerical calculations. We have shown that some parts of the nature of the magic square are appeared approximately in the Dirac flavor neutrino mass matrix and the magic squares prefer the normal mass ordering rather than the inverted mass ordering for the Dirac neutrinos. Moreover, we have mentioned about the recently reported a tension between T2K and NOvA experiments. It turned our that the T2K result is more favorable than the NOvA result in the context of the magic square.

Finally, we would like to comment about Eq.(\ref{Eq:magic_dirac}). In this paper, we have estimated the compatibility of the magic square with the real matrix in Eq.(\ref{Eq:magic_dirac}) instead of the complex Dirac mass matrix in Eq.(\ref{Eq:M-Dirac}). This approach may be enough as a first step of the study  about the relation between the magic square and the Dirac flavor neutrino mass matrix; however, if we estimate the compatibility of the magic square with the complex Dirac mass matrix in Eq.(\ref{Eq:M-Dirac}) by an appropriate method, the results may be modified. A detailed analysis of this topic will be found in our future study.

\vspace{3mm}







\end{document}